\documentclass[journal,comsoc]{IEEEtran}
\usepackage[T1]{fontenc}

\ifCLASSINFOpdf
\else
\fi
\usepackage{amsmath}
\usepackage{graphicx}        
\usepackage{subfigure}
\usepackage{graphicx}
\usepackage{array}
\usepackage{tabularx}
\usepackage{longtable}
\usepackage{booktabs}
\graphicspath{{figs/}}

\usepackage{pifont}

\usepackage{cite}

\usepackage[citecolor=black]{hyperref}
\usepackage{mfirstuc}
\usepackage{xcolor}

\newcommand{\tabincell}[2]{\begin{tabular}{@{}#1@{}}#2\end{tabular}}

\begin{document}
%
\title{Vision, Requirements, and Technology Trend of 6G\\ -How to Tackle the Challenges of System Coverage, Capacity, User Data-rate and Movement Speed}
%
%
%

\author{Shanzhi~Chen,
        Ying-Chang~Liang,
        Shaohui~Sun,
        Shaoli~Kang,
        Wenchi~Cheng
        and~Mugen~Peng

\thanks{This work was supported in part by the National Science Fund for Distinguished Young Scholars in China under grant 61425012, and the National Science Foundation Project in China under grant 61931005 and 61731017.}\thanks{Shanzhi Chen (Corresponding Author: chensz@datanggroup.cn) is with the State Key Laboratory of Wireless Mobile Communications, China Academy of Telecommunication Technology, Beijing 100191, China.}\thanks{Ying-Chang Liang (e-mail: liangyc@ieee.org) is with University of Electronic Science and Technology of China, Chengdu.}\thanks{Shaohui Sun (e-mail: sunshaohui@catt.cn) and Shaoli Kang (e-mail: kangshaoli@catt.cn) are with Datang Mobile Communication Equipment Co., Ltd, Beijing, China.}\thanks{Wenchi Cheng (e-mail: wccheng@xidian.edu.cn) is with Department of Telecommunication Engineering, Xidian University, Xi'an, China.}\thanks{Mugen Peng (e-mail: pmg@bupt.edu.cn) is with Beijing University of Posts and Telecommunications, Beijing, China.}}

\maketitle

\begin{abstract}
Since 5G new radio comes with none standalone (NSA) and standalone (SA) versions in 3GPP, research on 6G has been on schedule by academics and industries. Though 6G is supposed to have much higher capabilities than 5G, yet there is no clear description about what 6G is. In this paper, a comprehensive discussion on 6G is given based on the review of 5G developments, covering visions and requirements, technology trends and challenges, aiming at tackling the challenge of coverage, capacity, the user data rate and movement speed of mobile communication system. The vision of 6G is to fully support the development of a \textbf{\emph{Ubiquitous Intelligent Mobile Society}} with intelligent life and industries. Finally, the roadmap of 6G standard is suggested for the future.
\end{abstract}


%
\IEEEpeerreviewmaketitle

\section{Introduction}
\IEEEPARstart {M}{obile} communication systems upgrade to a new generation in every 10 years. In particular, research on new generation will formally start right after the earlier generation gets standardized. For the present fifth generation (5G) mobile communication system, the standard of none-standalone (NSA) version was finished at the end of 2017 and a standalone (SA) version was finalized in the middle of 2018 by the 3rd Generation Partnership Project (3GPP). Though further enhancements for 5G new radio (NR) are still going on, researches on the next generation, namely 6G, have been discussed by academics and industries. For example, in the European Union, some research projects such as Terapod and Terranova were sponsored by 5GPPP Phase 1 in 2017, then more projects were followed in Phase 2 and Phase 3 aiming at 6G innovations. In the US, following the joint university microelectronic project (JUMP) launched by Defense Advanced Research Projects Agency (DARPA) in 2016, the frequency band from 95GHz to 3THz were granted by Federal Communications Commission (FCC) for research on 6G. Also, in Chinese Communication Standardization Association (CCSA), two study items were launched in 2018, one focusing on the vision and requirements of 6G and the other on the key technologies for 6G. To better understand 6G and for the benefit to do research on 6G, based on the review on the present 5G, this paper tries to discuss a series of key issues on 6G, including vision and requirements, technology trend and challenges, as well as roadmap, etc.

The remaining part of the paper is organized as follows. In Section II, the progress of 5G development is reviewed, and the vision and requirements of 6G are discussed. In Section III, the key technologies of 6G are discussed. In Section IV, we highlight the roadmap of 6G development. And finally, in Section VI we give the conclusions.

\section{Vision And Requirements Of 6G}

\subsection{Review of 5G Development}

\noindent To review the process of 5G, the pace of the development is fast. For example, in 2013, different organizations were launched and various white papers were issued. In 2014, vision and requirements for 5G were defined by ITU. In 2015, key technologies were identified by academics and industries. In 2016, standardization was formally launched in 3GPP, and also technology trials were conducted by major official organizations. In 2017, the first version NSA standard was finished. In 2018, the second version SA standard was finalized. In 2019, 5G pre-commercial network is under development widely in the world.

The vision and requirements are defined by ITU: 5G should satisfy three typical scenarios and eight key performance indexes (KPIs) \cite{series2015imt}. Challenging technical indicators of three scenarios include Gbps data rate of enhanced mobile broadband (eMBB), millisecond (ms) air interface delay of ultra-reliable low latency communication (URLLC), and one million connections per square kilometer (1M/km$^{2}$) of massive machine type communication (mMTC). To satisfy these KPIs, a series of enabling technologies were proposed and discussed in standardizations and implemented in technology trials \cite{chen2014requirements}. Wireless technologies include massive MIMO, advanced coding and modulation, mmWave communication, non-orthogonal multiple access, ultra-dense networking (UDN), dual connectivity architecture, flexible frame structure, etc. Network technologies include network slicing, mobile edge computation (MEC), software defined network (SDN), network function virtualization (NFV), customized mobility, service based architecture, etc. Unfortunately, at the moment, there are two different aspects between definitions and actual realizations. One comes from the progress of 5G NR standardization, scenarios of eMBB and URLLC have been included in 5G NR NSA and SA version, while the mMTC scenario has not been agreed into NR yet being represented by narrow band internet of things (NB-IoT). The next comes from the progress of 5G industry development, scenario of eMBB is given higher priority and has developed much faster than the other two scenarios URLLC and mMTC. Reasons are difficulties occurring when pushing forward these IoT related scenarios, such as diversified, fragmented, and unclear requirements, etc. In conclusion, superior to former generations, 5G brings the groundbreaking aims of interconnection of everything (IoE) and breaks through the traditional layout of mobile communication. It has met a series of challenges from standardization and implementation, etc. Since it's difficult to solve all related problems in 5G stage, 5G can be regarded as just the starting phase of IoE. Therefore, research beyond 5G will aim to remove those differences between definition and realization and then surpassing existing vision and requirements.



Here we give further explanations. Just as the developing experiences from former generations, 2G is the best solution for voice mobile communication though voice was early defined in 1G, and similarly 4G is the best solution for mobile internet though mobile internet was early defined in 3G. There comes a magic word ''1G short, and 2G long, 3G short, and 4G long''. What will be the lifetime of 5G? The goal of 5G is to solve the interconnection of everything (IoE), but 5G vertical applications (internet of vehicles, industrial internet, etc.) is a very difficult and long-term process. IoE may thus only be the early idea from 5G technology experts, yet many issues are not clear at the moment, such as what will be the future actual demands? what will be the 5G pain points for industry application? what will be the industrial ecology? These key issues challenge the 5G's ability to satisfy IoE, which is actually what 6G needs to solve. It's rather like imperfect 3G fulfilled by 4G.

Now, the users volume of former generations (mainly 2G, 3G, and 4G terrestrial mobile communication networks) is about 70\% of the wordwide population. Due to the costs and technology limitations, those networks cover only about 20\% of the global land area, and less than 6\% of the earth's surface. For example, more than 80\% of the land area and 95\% of the sea area of China are not covered by terrestrial mobile communication networks. Since 5G is originally defined as a terrestrial mobile communication, it will also meet the above radio coverage limitation. That is to say, to make everything interconnected, future mobile communications should cover the entire surface area of the earth, including oceans, deserts, forests and the airspace.

\subsection{The Vision And Requirements of 6G}

\noindent From the above analysis on 5G developments, it can be foreseen that on one hand 6G should solve the limitations of 5G including at least system coverage and IoE. On the other hand, to meet the mobile communication demands of year 2030 and beyond, 6G should make the human society as a \textbf{\emph{Ubiquitous Intelligent Mobile Society}}. This becomes the vision of 6G.

Based on the vision and 5G development, 6G will be further upgraded and expanded to achieve ten to hundred times higher data rate, higher system capacity, higher spectrum efficiency, lower delay, as well as wider and deeper coverage, to support higher moving speed, to serve the interconnection of everything, and to fully support the development of \textbf{\emph{Ubiquitous Intelligent Mobile Society}} for intelligent life and industries. Following paragraphs will give detailed descriptions about foreseen requirements related to the vision.

\begin{figure}[htp]
  \centering
  \renewcommand{\figurename}{Figure}
  \includegraphics[width=8cm]{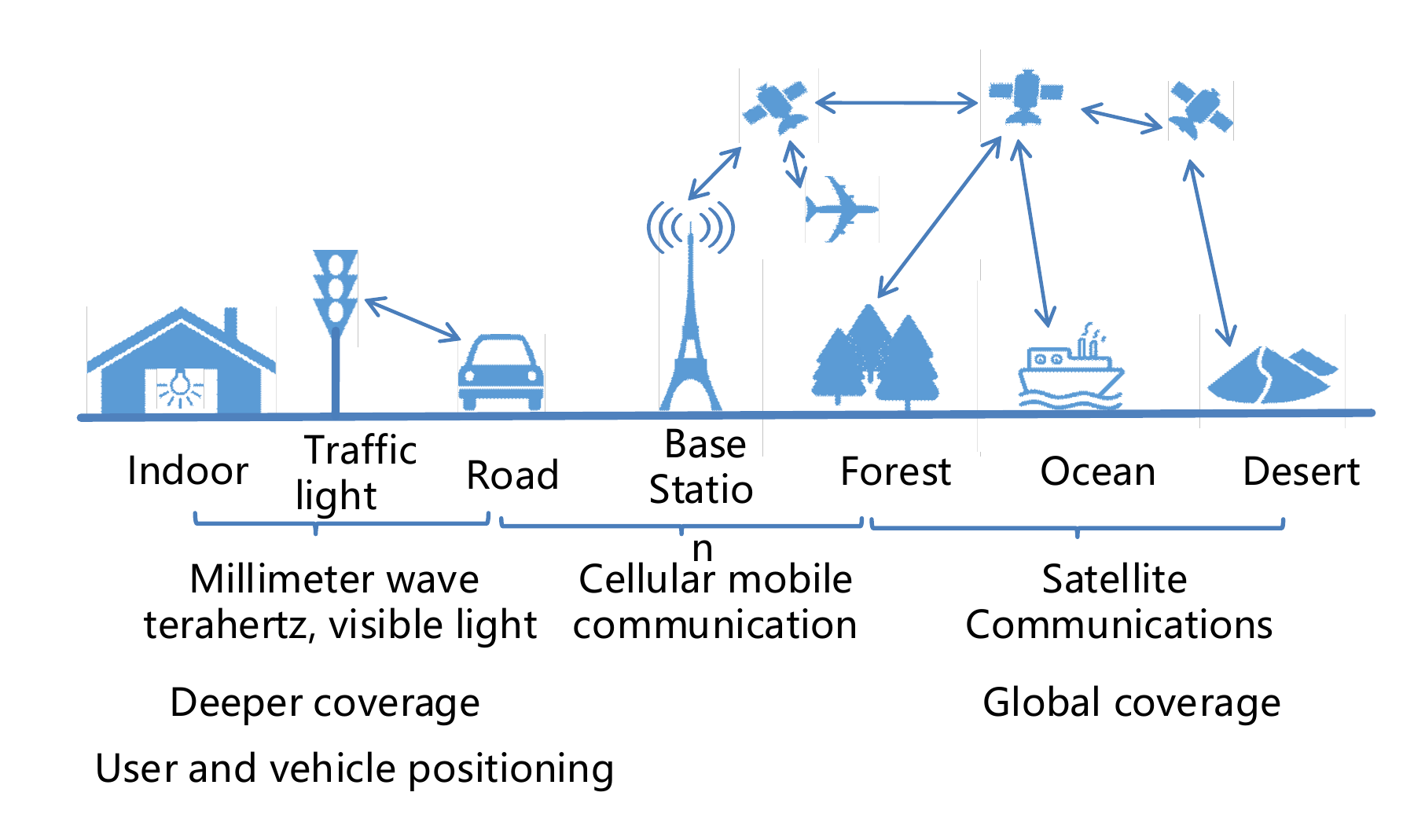}
  \caption{6G with global and deeper coverage.}
  \label{Fig:fig1}
\end{figure}

The first, 6G should be a ubiquitous and integrated network with broader and deeper coverage, including terrestrial communication, satellite communication, device to device communication in short range, etc. With the intelligent mobility management technology,  6G can serve in various environments such as airspace, land and sea, realizing global ubiquitous mobile broadband communication system, as shown in Fig. \ref{Fig:fig1}.

The second, 6G is expected to work on higher frequency to achieve wider bandwidth, such as mmWave, Terahertz, visible light, etc. Compared with 5G, 6G can promote the data rate up to ten to hundred times, supporting Tbps peak data rate and 10Gbps user experience data rate. In addition, 6G can use flexible frequency sharing technology to further enhance the frequency reuse efficiency.

The third, 6G is a personalized intelligent network. Combined with artificial intelligence technologies, 6G will realize virtualized personal mobile communication, with the network changing from a traditional function centralized type into a novel 3-Centralized type, i.e., user centralized, data centralized, and fully content centralized.

The fourth, 6G network will have endogenous security scheme or function security integrated design. By introducing trusty and safety mechanism, 6G has the capability of self-awareness, real-time dynamic analysis, and adaptive risk and confidence evaluation, all of which will help realizing cyber space security.

The fifth, 6G will merge computation, navigation and sensing with communications. For example, 6G will include not only satellite communication system, but also satellite navigation and positioning system, and even radar sensing system. 6G will adopt a more open architecture, with software defined core network and radio access network; 6G can realize fast and self-intelligent development and rapid dynamic deployment of network functions.

Finally, 6G can generate massive data through internet of everything; also, 6G can combine with novel technologies such as cloud computing, edge computing, artificial intelligence, block chain, etc; 6G can realize everything intelligence and group collective intelligence (swarm intelligence); and 6G can finally support a \textbf{\emph{Ubiquitous Intelligent Mobile Society}}.

\subsection{APPLICATION SCENARIOS AND CAPABILITIES OF 6G}

\noindent With the above expected vision and requirements of 6G, 6G will have much wider application scenarios than that of 5G. Scenarios include not only traditional eMBB, mMTC, URLLC scenarios, but also much upcoming ones, such as holographic communication, personal monitoring, drone taxi, internet of robots, wireless brain-computer interactions, etc. Similar to that virtual and augmented reality (AR/VR) is one of the most important application of 5G, holographic communication will be one of the most important application of high throughout requirement of 6G. It is expected that ten years later (2030~), media interaction will be developed from the current plane multimedia to high fidelity AR/VR, and finally to wireless holographic communication.

To show the 6G capabilities, Table \ref{Tab:table1} tries to give a comparison between 6G and 5G. Comparing original KPIs defined in 5G \cite{series2015imt}, 6G is expected to have 10 times enhancement on data rate, density and energy efficiency. 6G is expected to have about 3 times enhancement on mobility and spectrum efficiency; and 6G is expected to further decrease its latency to be below 1 millisecond. Also, for those newly defined KPIs, 6G will enhance coverage percentage from present 70\% to 99\%; 6G will improve reliability from present 99.9\% to 99.999\%; 6G will reduce the positioning error from the present meter level to centimeter level; and 6G will even improve the receiver sensitivity better than -130dBm, etc.

\begin{table}[!t]
	\renewcommand{\arraystretch}{1.3}
	\caption{ Possible capabilities of 6G in comparison with 5G}
	\label{Tab:table1}
	\centering
	\begin{tabular}{lllllll}
		\toprule
		{Major factors}&\tabincell{l}{6G}&\tabincell{l}{5G}\\
		\midrule
		Peak data rate&	\tabincell{l}{>100Gbps} & 10[20]Gbps  \\
		\specialrule{0em}{2pt}{2pt}
		\tabincell{l}{User experience \\ data rate } & >10Gbps & 1Gbps \\
		\specialrule{0em}{2pt}{2pt}
		\tabincell{l}{Traffic density} & >100Tbps/$km^{2}$ &	10Tbps/$km^{2}$\\
		\specialrule{0em}{2pt}{2pt}
		Connection density&	\tabincell{l} >10million/$km^{2}$ & \tabincell{l} 1million/$km^{2}$ \\
		\specialrule{0em}{2pt}{2pt}
		Delay&	<1ms & ms level\\
		\specialrule{0em}{2pt}{2pt}
		Mobility &	>1000km/h &	\tabincell{l} 350km/h\\
		\specialrule{0em}{2pt}{2pt}
		Spectrum efficiency &\tabincell{l} >3x relative to 5G  & 3$\sim$5x relative to 4G\\
		\specialrule{0em}{2pt}{2pt}
		Energy efficiency & \tabincell{l} >10x relative to 5G & 1000x relative to 4G\\
		\specialrule{0em}{2pt}{2pt}
		Coverage percent & \tabincell{l} >99\%  & about 70\% \\
		\specialrule{0em}{2pt}{2pt}
		\tabincell{l}{Reliability}& \tabincell{l}> 99.999\%  & \tabincell{l}about 99.9\% \\
		\specialrule{0em}{2pt}{2pt}
		Positioning precision & centimeter level & meter level\\
		\specialrule{0em}{2pt}{2pt}
		Receiver sensitivity & <-130dBm & about -120dBm \\
		\bottomrule
	\end{tabular}
\end{table}

\section{Technology Trend And Challenges}
	
\begin{figure*}
  \centering
  \renewcommand{\figurename}{Figure}
  \includegraphics[width=16cm]{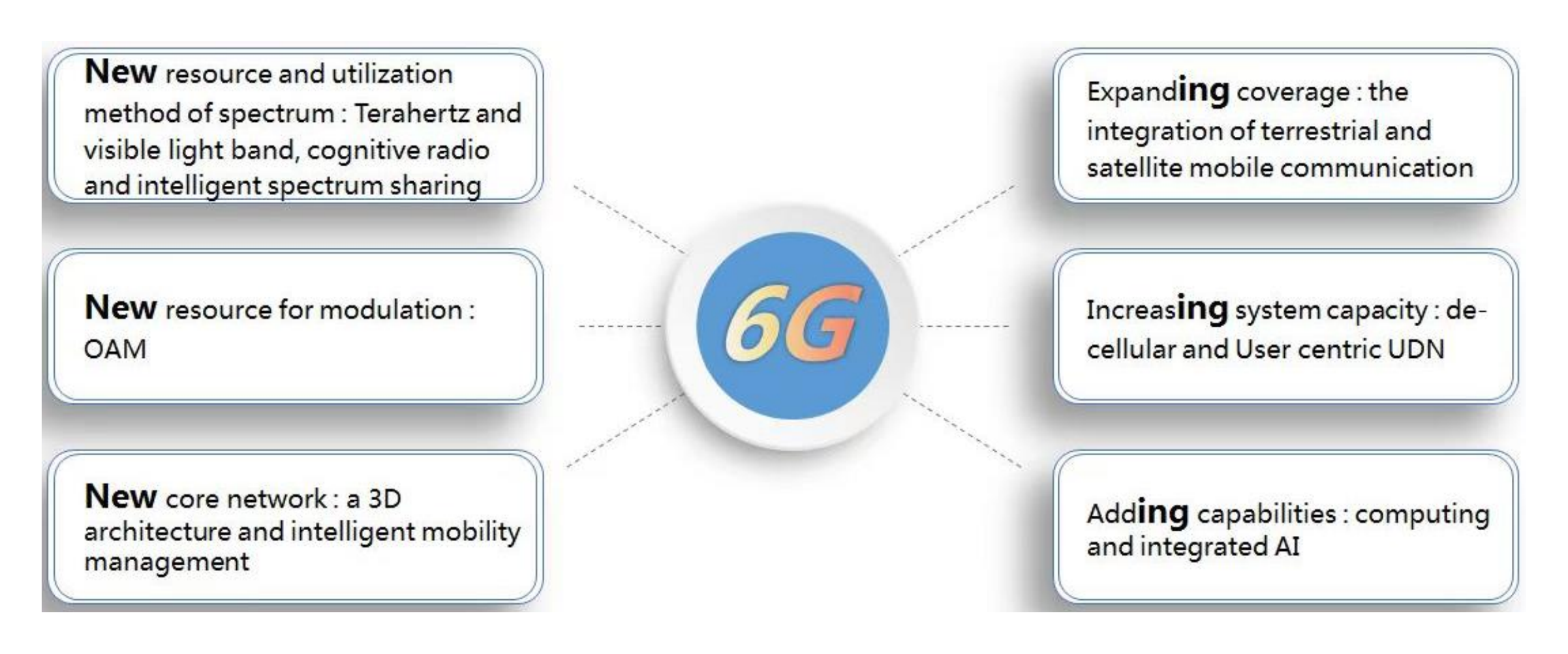}
  \caption{Key technologies and new methods for 6G and Beyond.}\label{Fig:fig2}
\end{figure*}

\noindent The challenges of 6G include system coverage and capacity, user data rate and movement speed, spectrum and energy efficiency. To follow up listed challenges, 6G can be represented using ''\textbf6'' key aspects, including three ''\textbf{New}'' resources and methods and three ''\textbf{ing}''-related key technologies in \color{black}{Fig. \ref{Fig:fig2}}.

The realistic way to solve the problem of global coverage and higher-speed user movement is to adopt satellite mobile communication system. Compared with terrestrial mobile communication system, satellite mobile communication system can achieve wider coverage (such as over-the-sea, forest, and desert) at low cost. When moving speed of a mobile terminal is beyond Mach level, it is difficult to be supported by the terrestrial mobile communication system; yet it is easy to be supported by satellite mobile communications. Since 6G will integrate land wireless mobile communication, medium and low orbit satellite mobile communication, and also short range direct communication technologies into one mobile system, there are thus challenges to design a new core network architecture and mobility management.

There are four methods to enhance 6G system capacity. The first one is to add more spectrum bandwidth. The second one is to increase the spectrum efficiency of the air interface (bits per second per Hertz) by adopting better modulation and channel coding, or new modulation resource. The third one is spectrum reuse by reducing the cell size. The fourth one is new spectrum utilization, such as flexible spectrum sharing technology. Ways to offer user data rate mainly come from the first two methods in the list from one to four for system capacity.

In addition, 6G will develop the new air interface that enables multiple heterogeneous wireless transmission accesses, and 6G will merge computation, navigation and sensing, both of which will make 6 Ga much more complex system, requiring AI technology and computing capability to empower 6G.

In-depth analysis of some of these listed technologies are elaborated below.

\subsection{Expending Coverage: The Integration Of Terrestrial And Satellite Mobile Communication}
	
\noindent The prospect of future global communication networks is to offer fast and integrated service to ubiquitous users with on-demand personality at any time. Satellite communication has advantages to support system coverage and user moving speed, and it will play an important role in the 6G infrastructure.

According to the height of orbit, satellites can be classified as the low-earth orbit (LEO) satellite with height about 500 to 2000km, the medium-earth orbit (MEO) satellite with height about 2000 to 36000km, and the geostationary (GEO) satellite with height about 36000km.

Conventional terrestrial and satellite mobile communication systems were established with separate standards in ITU. They have different characteristics as shown in Table \ref{Tab:table2}. For example, at 3G stage, three kinds of CDMA systems, CDMA2000, WCDMA and TD-SCDMA were approved for terrestrial mobile communication, while eight systems, SRI-A to SRI-H were approved for satellite mobile communication. Similarly, at 4G stage, following LTEs were approved as 4G terrestrial mobile communication standards in WRC-12, SAT-OFDM and BMSat were also approved by ITU as 4G satellite mobile communication standards \cite{detailedspecifi} in September 2013.

\begin{table}[!t]
 \renewcommand{\arraystretch}{1.3}
 \caption{Differences between terrestrial and satellite mobile communication systems}
 \label{Tab:table2}
 \centering
 \begin{tabular}{lllllll}
\toprule
    {Major factors}&\tabincell{l}{Satellite mobile \\ communication}&\tabincell{l}{Terrestrial mobile \\ communication}\\
\midrule
    Link type&	\tabincell{l}{Including service link \\and feeder link} & Service link only  \\
    \specialrule{0em}{2pt}{2pt}
    \tabincell{l}{Transmission \\ distance} & Above 600km & About 1 km \\
    \specialrule{0em}{2pt}{2pt}
    \tabincell{l}{Transmission \\delay} & Tens to hundreds millisecond&	us to ms level\\
    \specialrule{0em}{2pt}{2pt}
    Frequency&	\tabincell{l}{Usually high \\frequency band \\like Ku, Ka, Q, etc} & \tabincell{l}{Usually low \\frequency band \\below 6GHz}\\
    \specialrule{0em}{2pt}{2pt}
    Path loss&	Often above 180 dB & Usually within 140 dB\\
    \specialrule{0em}{2pt}{2pt}
    Doppler shift &	Up to several hundred KHz&	\tabincell{l}{Usually within \\KHz level}\\
    \specialrule{0em}{2pt}{2pt}
    Cell Radius&	Often several hundred km&	Usually300-500 m\\
    \specialrule{0em}{2pt}{2pt}
    Mobility & \tabincell{l}{Frequently happen, including \\inter-beams (inter-cells), \\inter-satellites, inter-gateways.} & Inter-cells only\\
    \specialrule{0em}{2pt}{2pt}

    \tabincell{l}{Requirements \\for PAPR}&	\tabincell{l}{Strictly low PAPR to \\avoid power waste} & Low PAPR\\
    \specialrule{0em}{2pt}{2pt}
    \tabincell{l}{Frequency \\reuse}& \tabincell{l}{Different frequencies with \\reuse factor 4,8, etc.} & \tabincell{l}{Same frequency with \\reuse factor 1.}\\
 \bottomrule
\end{tabular}
\end{table}

To go beyond 5G and to integrate satellite mobile communication into terrestrial mobile communication, 3GPP started a study item (SI) Study on NR to support Non Terrestrial Networks (NTN) in RAN1 of Release 15. The SI discusses channel modeling and design constraints and analyzes influences to NR. A technical report TR38.811 \emph{Study on NR to support Non Terrestrial Networks} was formed eventually in 2018. In consecutive release, a further SI  \emph{Solutions for NR to support Non Terrestrial Networks} was conducted in 3GPP, trying to explore technique solutions.

For satellite communication system, usually the LEO system is given the widest attention and regarded as the hottest topic in recent years. Due to lower orbit height, compared with MEO or GEO, LEO has priorities like shorter transmission delay, smaller path loss, etc. Furthermore, thanks to the state of the art satellite technologies, multiple satellites can be launched together by one rocket, the total cost for satellite system is much reduced. With hundreds or thousands of satellites located at dozens of orbits above the earth, LEO system can thus authentically realize the global coverage and more efficiency by frequency reuse. In addition, advanced mobile technologies provide technique guarantee for the success of LEO system, such as cellular communication, multiple access, spot beam, frequency reuse, etc. Therefore, LEO system is deemed as the most prospective satellite mobile communication system.

However, there also exists some inherent disadvantages in LEO system, for example, the large number of satellites induce higher complexity in control, operation, and management. Furthermore, from the communication aspect, there meets integrating challenges for LEO and terrestrial systems. Major challenges are listed in the following paragraphs.

\noindent\textbf{Doppler shift and Doppler variation:} Since a LEO satellite moves much faster than the earth rotation, it brings evident Doppler shift and Doppler shift variation in communication. Taking the orbit height 600km as an example, the Doppler shift reaches up to 480KHz at Ka-band 20GHz carrier, which will seriously affect communication process like synchronization, random access, signal measurement, signal detection, etc.

\noindent\textbf{Large transmission delay:} Since the satellite transmission distance is much longer than that of terrestrial transmission, the LEO signal transmission delay and the path loss are much higher than that of terrestrial system, resulting in non-negligible differences in designs on aspects like waveform, modulation, channel coding, HARQ, MAC, RLC, etc. For example, due to power limitation from the satellite, the waveform and modulation should be redesigned to make the PAPR as low as possible. Due to larger delay for signal transmission, HARQ should be redesigned to adapt to the satellite transmission.

\noindent\textbf{Transmission Technology for Inter-satellite Link (ISL):} Usually there are two types of communication modes from satellite, one is the bent-pipe mode, and the other is the regenerative mode. For the regenerate mode, a satellite has the functions of detection, demodulation and modulation on communication signals. Even the routing function can be included in satellite, satellites can thus communicate with each other directly via the inter-satellite link (ISL). ISL will stimulate research on innovative transmission technology, for example, optical communication and Terahertz (300GHz-10THz) radio communication are important research directions in recent years. When Laser and Terahertz are used for ISL, they can provide high date rates up to several Gbps, to ensure the requirements of capacity and throughput. ISL transceiver has advantages of small size, light weight, low power consumption, strong ability to resist electromagnetic interference, and high security and confidentiality. However, due to the narrow beam in transmission and high moving speed, it is tough for ISL beam searching, targeting, capturing and tracking. In addition, since an ISL has long transmission distance, the signal attenuation is heavy, resulting in challenges on receiver sensitivity, signal detection quality, and receiving performance.

\subsection{New Resource And Utilization Method Of Spectrum}
	
\subsubsection{Terahertz And Visible Light Band}

\noindent The utilization of still-unregulated high-frequency band can alleviate the spectrum scarcity and capacity limitations of current wireless communication systems. The Terahertz (THz) band (from 0.1 THz to 10 THz) and visible light band (from 400 THz to 800 THz) are recognized as two ultra-wide spectral bands to enable ultra-high-rate communications. The corresponding THz communications and visible light communications (VLC) are two promising technologies for 6G and beyond\cite{pathak2015visible}.

On the one hand, THz communications rely on advanced THz devices to transfer information at THz band. Currently there are two kinds of THz communication systems, including solid state THz communication system based on frequency mixing mechanism, and a spatial direct modulation THz communication system which modulates baseband signals directly into a continuous THz carrier wave. THz communications have diverse application scenarios, such as the indoor wireless mobile networks supporting holographic video conference and virtual reality, nanoscale communication networks for health monitoring via nano-machines, and space communication networks like inter-satellite communication and near-space communication.

Although THz communications have obvious advantages for 6G and beyond applications, there are still many technical challenges to be solved before practical deployment.

\noindent\textbf{High-frequency hardware components:} For solid state THz communication systems, it is difficult to design efficient radio-frequency (RF) circuits such as THz mixers, THz oscillators, THz amplifiers and THz antennas. Specifically, it is a great challenge to design ultra-broadband THz antennas with high gain and fast beam scanning function. Physically, the low noise design for such high bandwidth super-heterodyne transceiver is also a worldwide challenge. For direct modulation THz communication systems, one determining factor is the performance of THz modulator. Specifically, it is desirable to design amplitude THz modulator with high modulation speed and depth, as well as phase modulator with large scale and linear phase shifts. The advanced materials and device structures such as GaN-HEMT and Graphene are promising for THz modulator design\cite{hasan2016graphene}.

\noindent \textbf{THz communication channel modelling and estimation:} To implement efficient wireless communication systems in THz band, it is imperative to establish accurate channel models for THz communications. Different from lower-frequency channels, the THz channel modelling needs to consider several unique factors, including the highly frequency-selective path-loss due to the absorption loss of oxygen and water-vapor molecule, mutual coupling effect and near-field effect together with spatial non-stationarity over ultra-massive antenna arrays.  It is thus desirable to have efficient channel estimation schemes with low computing load for THz communication systems with ultra-massive antenna arrays.

\noindent\textbf{Directional networking for THz wireless communication:} Through the beams of THz band are relatively narrow compared with lower frequency band. The directional antenna is still preferable for THz communication networking, since it is capable to concentrate the THz wave energy in a particular direction to support longer-distance communication, and to reduce interference towards the neighboring nodes. The challenges for THz-band directional networking include efficient neighbor-discovery algorithms under time asynchronous system, topology control algorithms with optimized trade-off between node degree and jump stretch, and multiple-access-control (MAC) protocol with higher access capacity and lower resource consumption.

On the other hand, VLC are realized by transmitter performing intensity modulation (IM) on devices such as light-emitting diodes (LEDs). The VLC receiver performs direct detection (DD) on the received light by a silicon photodiode. VLC can achieve gigabits-per-second transmission with commercial LEDs through downlink. VLC have a variety of applications such as indoor high-speed data links for personal area networks, outdoor vehicle-to-vehicle and vehicle-to-infrastructure communication, and RF not friendly environments like mines and underwater networks \cite{pathak2015visible}

Despite VLC have inherent advantages such as ultra-wide frequency band, high energy efficiency, low cost and better safety, compared to traditional communication systems, VLC still faces following listed technical challenges:

\noindent\textbf{Bandwidth Enhancement:} Commercial LEDs typically have very limited modulation bandwidth and slow modulation response, which directly limit the achievable data rate of VLC systems. The LED bandwidth can be enhanced by various methods such as blue filtering, pulse shaping and subcarrier equalization. It is desirable to develop more advanced transmitter pre-equalization and receiver post-equalization techniques to address the modulation-bandwidth limitation of VLC systems.

\noindent\textbf{Non-linearity compensation:} The orthogonal frequency-division multiplexing (OFDM) is widely considered as the most powerful modulation technique for high-speed VLC. The relationship between the current through LED and the emitted light of a LED is inherently nonlinear. Hence, the LED's limited dynamic range and the high peak-to-average- power ratio constitute the main challenges for OFDM-based VLC systems. Sophisticated pre-distortion and post-distortion methods are expected for OFDM to counteract the effect of the nonlinearities.

\noindent\textbf{Multiple-input multiple-output (MIMO) techniques:} MIMO techniques are alluring to enhance the VLC rate, but MIMO systems are difficult to realize in VLC. First, the diversity gains are limited due to very similar paths between the transmitter and the receiver. Second, the design of VLC MIMO receiver is difficult. A non-imaging receiver requires accurate transmitter-receiver alignment, while an imaging receiver suffers from lower sampling rate of imaging sensor. Interference-mitigation techniques in VLC MIMO need to be further studied, such as multiuser precoding and coordinated-beamforming \cite{ma2018coordinated}.

\subsubsection{Cognitive Radio And Intelligent Spectrum Sharing}

\noindent Although new ultra-high-frequency bands from THz and visible light will be explored in 6G and beyond, the available spectrum resource especially the lower-frequency band with established coverage, is still scarce in terms of exponential growth of wireless data traffic. In 5G and pervious system, the dedicated spectrum allocation makes the spectrum resource fully occupied, and the utilization rate is low. Hence, efficient spectrum managements are still critical schemes to improve the spectrum efficiency of 6G communication systems. Cognitive radio (CR) techniques allow multiple wireless systems to share the same spectrum by utilizing spectrum sensing and interference management techniques. In particular, the spectrum sharing techniques enable secondary systems to utilize the unlicensed or underutilized spectrum temporally and geographically, which will significantly enhance the spectral efficiency \cite{zhang2017survey}.

As one of the most recent evolutions of CR and spectrum sharing, symbiotic radio (SR) is a new promising technique for 6G communication systems. By using the technique, multiple subsystems of a heterogeneous wireless communication system intelligently cooperate with each other to realize mutually beneficial transmission and efficient resource sharing. The technique further enhances overall efficiency of a CR system significantly. One typical example is an SR network based on ambient backscatter communications \cite{zhang2018constellation}, in which the IoT devices transmit information to their destinations by passively reflecting the signals received from the primary transmitter. The IoT transmission thus shares the same spectrum, signal source, and infrastructure of the primary system.

Despite the extensively studied conventional CR and spectrum sharing techniques, there are still critical challenges for new CR-inspired technologies including intelligent SR, efficient dynamic spectrum access, and intelligent spectrum sharing.

\noindent\textbf{Intelligent dynamic spectrum access:} Complete information about the network state is typically required in order to obtain optimal dynamic-spectrum-access (DSA) policies. However, existing DSA protocols cannot effectively adapt to more complex real-world models. There are studies applying deep reinforcement learning for designing distributed DSA algorithms to learn the hidden patterns of the primary systems \cite{zhang2019deep}. In order for the DSA algorithms applicable to complex spectrum sharing environments, new model-free distributed learning algorithms with lower computing overhead are highly desirable.

\noindent\textbf{Intelligent spectrum sharing:} A variety of spectrum sharing technologies can be used in 6G and beyond, such as device-to-device communication, in-band full-duplex communication, non-orthogonal multiple access, and spectrum sharing in unlicensed spectrum. To manage massive connections in 6G applications, distributed and efficient interference avoidance/mitigation techniques are required to enhance the system performance. The block chain and deep-learning technologies were initially shown to be efficient approaches for flexible spectrum sharing in \cite{weiss2019application}. Therefore, new framework and mechanisms for intelligent spectrum sharing need to be developed.

\noindent\textbf{Intelligent Symbiotic Radio:} SR is a new wireless communication paradigm, which aims to enhance the performance of the whole communication system via intelligent inter-subsystem cooperation. The subsystems may have different symbiotic mechanisms such as parasitism, commensalism and mutualism. Under each symbiotic mechanism, it is desirable to investigate the theoretical limits of fundamental information, and transmission theory with interference cancellation/mitigation/utilization. However, the research on SR is at the infant stage, and many open issues need to be addressed. For example, for large SR networks, the wireless access and multi-dimensional resource allocation are also challenging, which need to be addressed via techniques such as artificial intelligence and big data analysis.

\subsection{New Resource For Modulation: OAM}

Along with the fast development from 1G to 4G, different orthogonal resources such as frequency, time, code, space are extensively used. Although the non-orthogonal multiple access was discussed to be promoted in 5G,it is more demanded and attractive to explore a new resource for significant spectrum efficiency, user rate, and capacity enhancement in 5G-beyond and 6G wireless communications.


The electromagnetic wave has not only linear momentum, which is the foundation for the current wireless communications based on traditional plane-electromagnetic (PE) waves; there exists also angular momentum, which is being focused recently. OAM, as a kind of wave-front with helical phase, can be used as a new resource for modulation in wireless communications \cite{yang2018mode}. Spacing OAM has a great number of topological charges, that is, OAM-modes \cite{thide2007utilization}. Different OAM-modes are orthogonal with each other and can be multiplexed/demultiplexed together, bringing a new way for capacity enhancement in 6G and Beyond wireless communications networks.


\begin{figure*}
  \centering
  \subfigure[Radiation pattern of OAM-mode 0]{
    \label{fig:subfig:a} 
    \includegraphics[width=5cm]{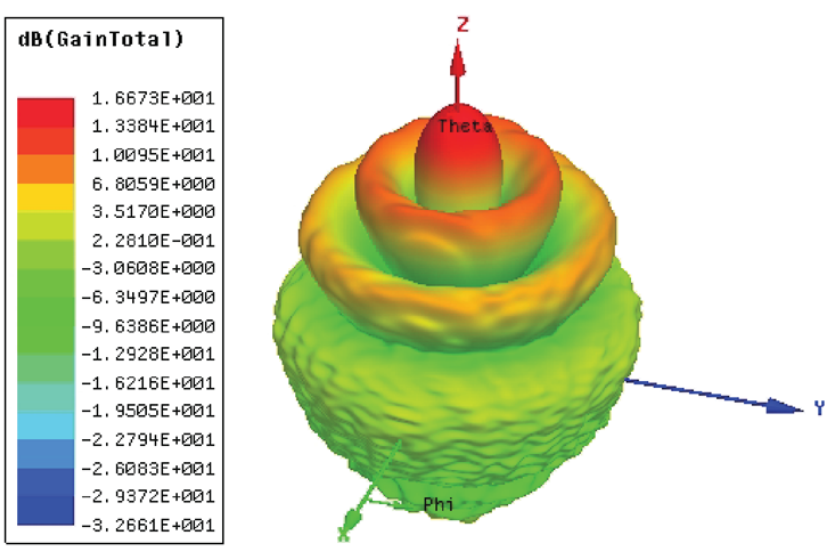}}
   \subfigure[Radiation pattern of OAM-mode 1]{
    \label{fig:subfig:b} 
    \includegraphics[width=5cm]{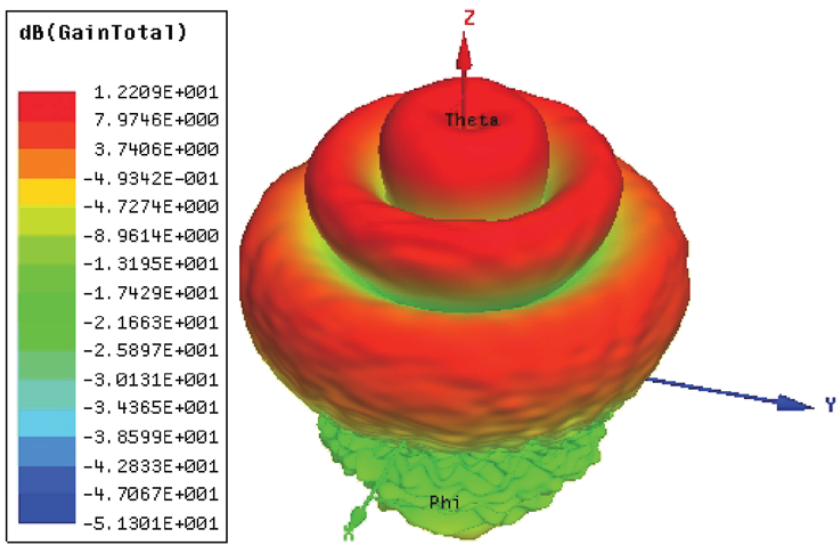}}
   \subfigure[Radiation pattern of OAM-mode 2]{
    \label{fig:subfig:c} 
    \includegraphics[width=5cm]{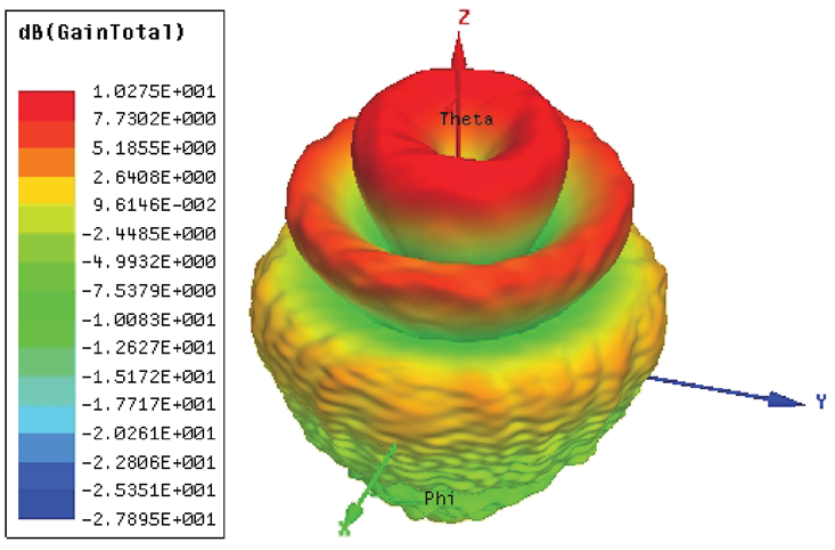}}
  \hspace{0.0in}
  \subfigure[Field intensity of OAM-mode 0]{
    \label{fig:subfig:d} 
    \includegraphics[width=5cm]{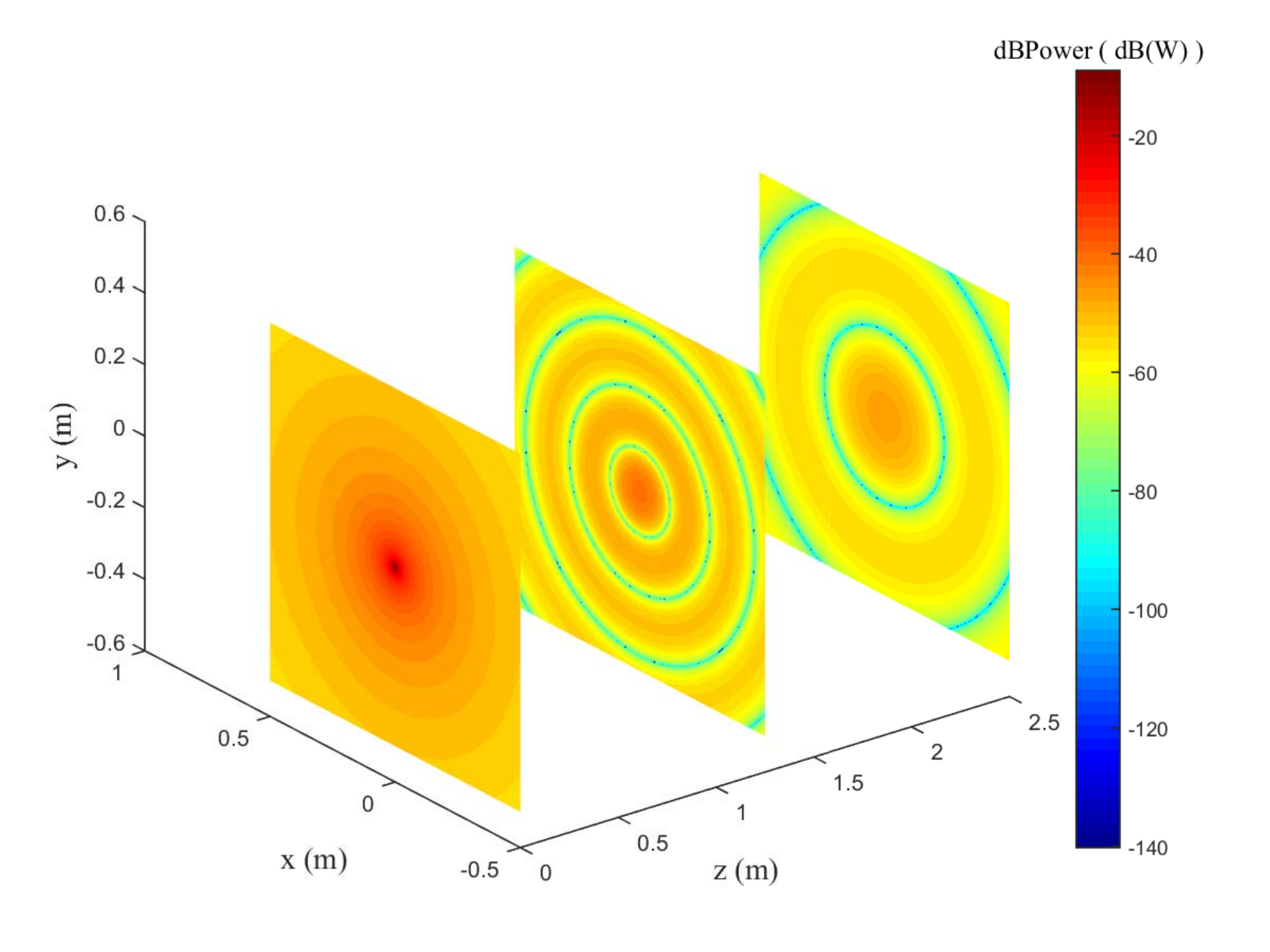}}
  \subfigure[Field intensity of OAM-mode 1]{
    \label{fig:subfig:e} 
    \includegraphics[width=5cm]{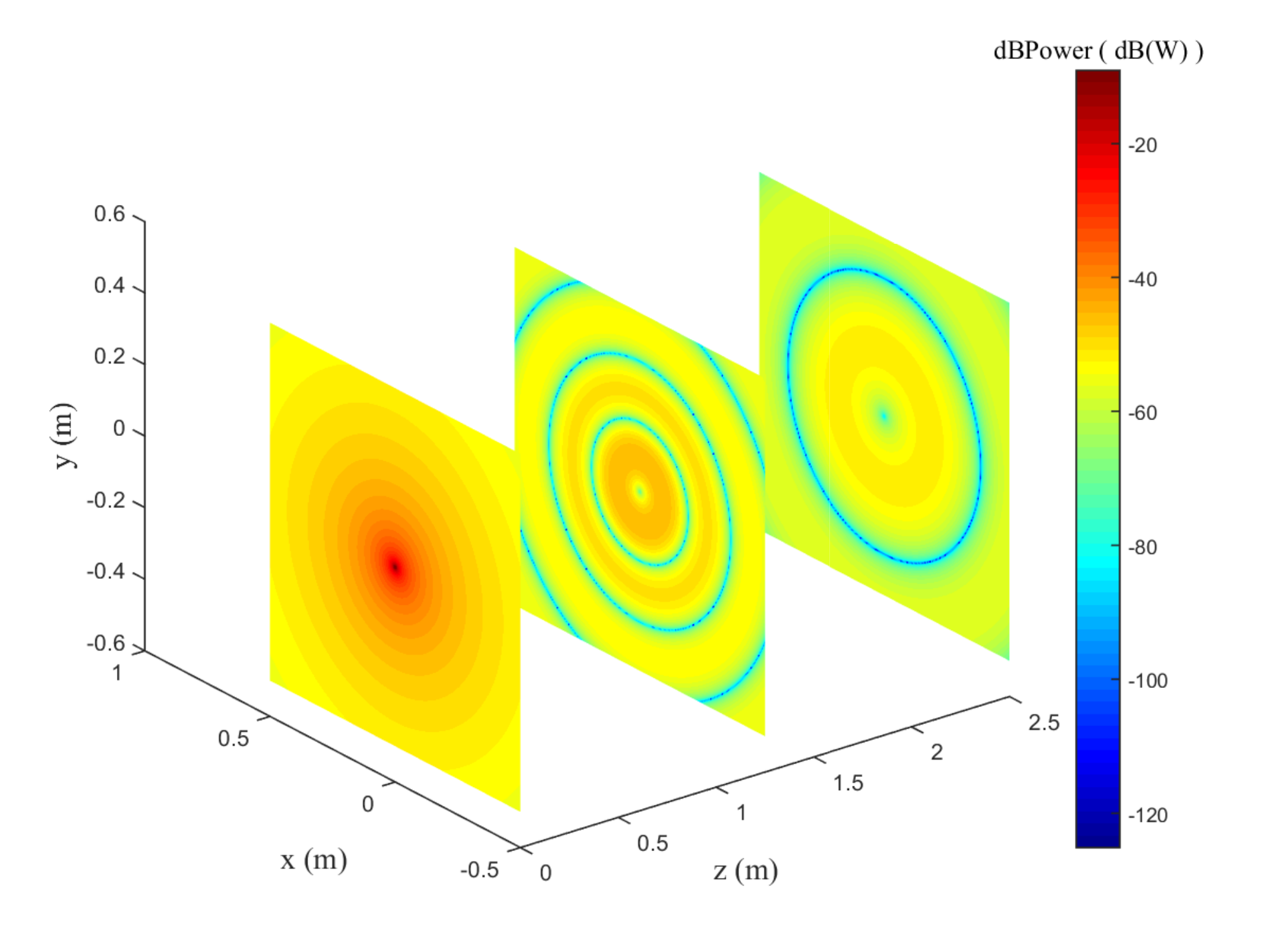}}
   \subfigure[Field intensity of OAM-mode 2]{
    \label{fig:subfig:f} 
    \includegraphics[width=5cm]{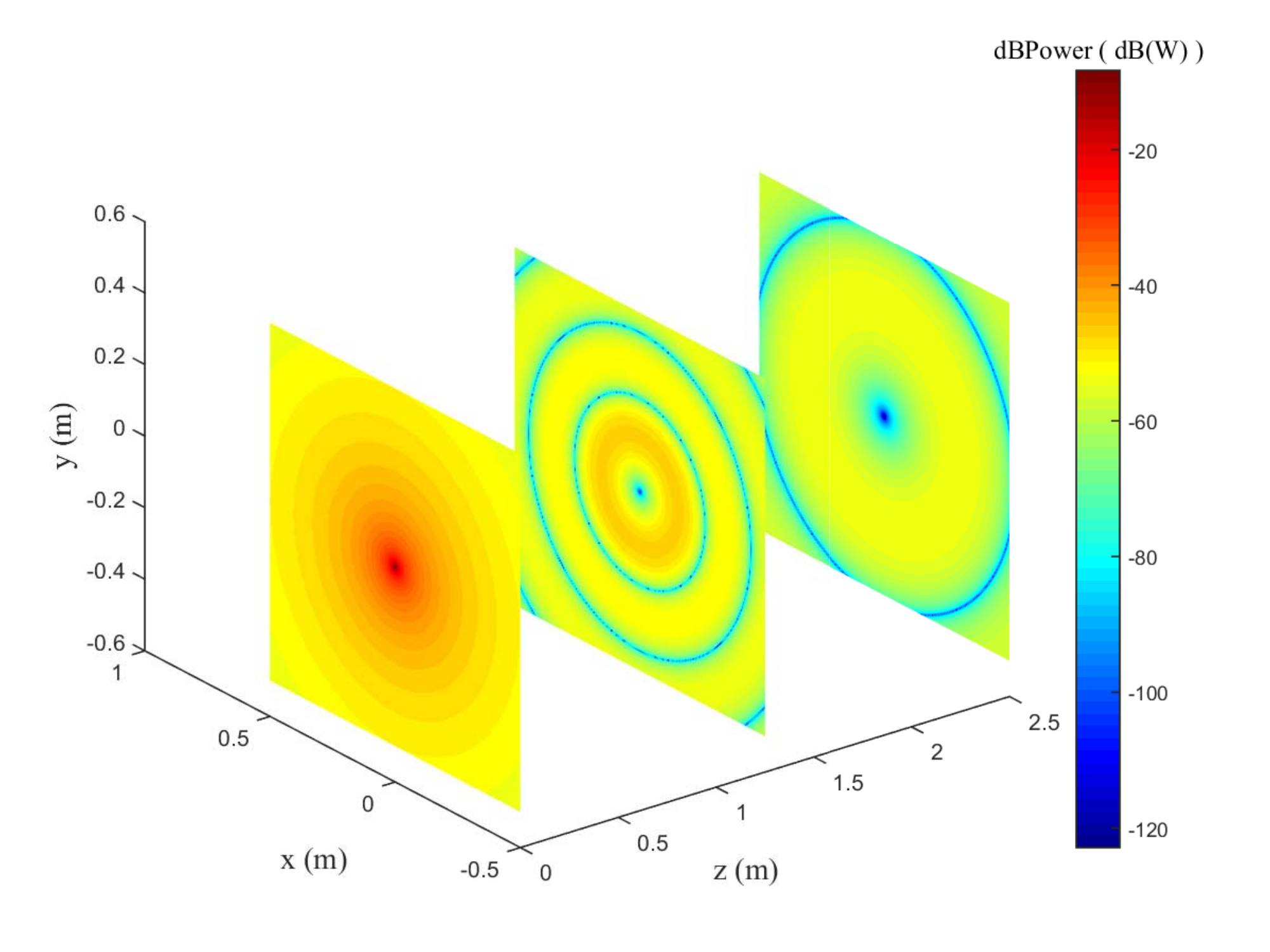}}
  \caption{Radiation pattern and field intensity for OAM waves with OAM-modes 0, 1, and 2.}
  \label{Fig:fig3} 
\end{figure*}

Fig. \ref{Fig:fig3} (a-c) shows the radiation pattern of OAM-modes 0, 1, and 2. Different data streams can be modulated on different OAM-modes. Fig. \ref{Fig:fig3} (d-f) plots the field intensity for different distances corresponding to different OAM-modes. The OAM waves are centrally hollow except OAM-mode 0, which is in fact the traditional PE wave. The central hollow pattern increases as the order of OAM-mode increases, which indicates that the degree of divergence increases as the order of OAM-mode increases.

Although it is intriguing that OAM can greatly increase the capacity and enhance other key metrics for wireless communications, three critical problems need to be solved before its practical implementation:

\noindent\textbf{Beam-divergence:}  OAM beams are divergent, which limits the efficient reception and long-distance propagation in OAM based wireless communication systems. It is desired that OAM beams are convergent for efficient transmission. There exist tentative schemes for converging OAM beams such as antenna based anti-divergence. The antenna based anti-divergence schemes employ the parabolic antennas or the lens antennas to converge OAM beams \cite{cheng2018orbital}. However, with the decreasing on degree of divergence, OAM beams, after antenna converging, are still with central hollow structure after relatively long-distance transmission.

\noindent\textbf{Transmission:} It is now commonly accepted that OAM can be used for line-of-sight (LoS) environments such as microwave backhaul scenarios and indoor massive data scenarios. For non-LoS (NLoS) environments, since the wave-front of OAM-modes changes with reflection and refraction of OAM waves, it is very difficult to model the OAM transmissions in fading scenarios. How to facilitate OAM based transmissions is an open problem for NLoS environments.

\noindent\textbf{Transceiver-misalignment:} Due to the phase sensitivity of OAM waves, current detection schemes are generally proposed for the transceiver-coaxial scenarios. However, it is usually the case that the transmitter and receiver are misaligned with each other in practical wireless communications scenarios. The transceiver-misalignment makes it challenging for distinguishing multiple OAM-modes at the receiver. Some phases-turbulence compensation schemes have been proposed for OAM based transmission in transceiver-misalignment scenarios.

\subsection{Increasing System Capacity: De-Cellular And UUDN}

\noindent As mentioned in the beginning of this section, to enhance system capacity, typical methods are introduction to more spectrum bandwidth resource and reuse of efficient spectrum by reducing the cell size, that is cell split. 6G is expected to work on higher frequency band to achieve wider bandwidth, such as mmWave, Terahertz, visible light, etc. Compared with 5G, wireless coverage range of 6G base-station (BS) or access point (AP) is thus getting even smaller. But it is impossible for the cell to split indefinitely, which approach to the limit of the system capacity improvement.

In the traditional cellular network, a single BS serves multiple users. The handover control occurs when a user cross different BSs. When the number of APs approaches to or even exceeds the number of user terminals as soon as the cell size and transmit distance become smaller and smaller, the cellular network architecture becomes ineffective. In this case, the de-cellular network architecture was proposed \cite{chen2016user}. This network will organize a dynamic APs Group (APG) to serve each user seamlessly without user's involvement, and let user feel like a network always following him. In this case, the network shall intelligently recognize the user's wireless communication environments, and then flexibly organize the required APG and resource to serve the user. User-Centric Ultra-Dense Networks (UUDN) was proposed in \cite{chen2016user} correspondingly, which is a wireless network with comparable densities of access point (AP) and the density of users. The proposal introduces the philosophy of network serving user.

The de-cellular and UUDN are suitable for the higher frequency band, such as mmWave, Terahertz, visible light band, yet there are several challenges:

\noindent\textbf{Network architecture:} How to design a network architecture to enable dynamically cooperate and transmit between APs, to match user's service requirement, to support continuous communication during its movement, and to enhance the spectrum efficiency and user experience. That is how to realize \emph{Dynamic network serving user}.

\noindent\textbf{Mobility Management:} While a user is moving, its AP group will be dynamically adjusted or refreshed to support its movement. The Dynamic APs Grouping method (DAPGing) was proposed in \cite{chen2016user} and need to be much further explored.

\noindent\textbf{Interference management:} Interference becomes more complex in the de-cellular network and UUDN. Dynamic ways such as Artificial Intelligence (AI) needs to be introduced to solve such complex problem.

\noindent\textbf{Resource Management:} The considerations of resource allocation should be based on users and a corresponding AP group, not based on single cell in traditional way.

\noindent\textbf{Security Management:} A new issue is the security mechanism among APs and AP groups. The trusted authentication and secure transmission should be built among APs, user to AP and AP groups. Because the AP group serving a user is generated dynamically, it is necessary to identify the security mechanism of the AP group when an AP joins or exits.

\subsection{Adding Capability: Computing And Integrated Artificial Intelligence}
	
\noindent To realize the vision of 6G, current wireless networks should be changed and revolutionized from the traditional function centralized type into a novel user centralized, content centralized, and data centralized (3C) type. Thus, adding computing and AI capabilities to the wireless networks is essential. In particular, to enhance transmit quality and decrease computing complexity for supporting ultra-wide spectral band in physical layer, to converge computing, communications, and data cache efficiently for achieving diverse performance, and to make the network architecture adaptive to ubiquitous application scenarios, AI is a trump card to extend the coverage, meet diverse requirements of massive applications, and improve performance significantly in 6G. Empowered by modern AI technologies, 6G can interconnect a massive number of devices from various environments, such as in the air, in the space, on the earth, and over the sea, and provide them unprecedented quality of service. In terms of applications, using AI for intelligent learning, inference, and decision making are just a few examples in 6G \cite{letaief2019roadmap}.

Although there are significant benefits, it is still challenging to achieve the real deployment of computing and AI capabilities in 6G. Specifically, the challenges are mainly as follows:

\noindent\textbf{Collaborative Intelligence and Computing among Cloud, Edge and Terminal:} The first key question for 6G is ''where to add the computing and AI capabilities, cloud or around edge?'' When adding the computing and AI capabilities to the cloud, the resource-limited end-user devices can enjoy the benefits of rich computing power and storage in the cloud and edge; while posing heavy load on the backbone networks the load may result in the unpredictable transmission latency. On the other hand, when deploying the computing and AI capabilities on the edge, the load on the backbone networks can be alleviated and the transmission latency can be significantly reduced. However, the end-user devices may face to large computation latency due to limited computing power and storage on the edge if the MEC scheduling policy is poor. As a result, collaborative intelligence and computing among cloud, edge and terminal should be revealed with highest priority.

\noindent\textbf{Heterogeneity and Dynamical Convergence:} 6G is highly heterogeneous in terms of the device type, the occupied frequency band, the infrastructure type, and service type. Such heterogeneity makes it prohibitively difficult and expensive to intelligently coordinate the communication, data caching, and computing resources. For example, the fog-computing-based radio access network architecture has been proposed to make communication, data caching and computing converge efficiently through flexibly assigning communication, data caching and computing functions in different entities. Meanwhile, the intent-based radio access network has been presented to make the network architecture flexible and simple through software-definition and re-configuration for the ubiquitous intelligent mobile society with low cost. Therefore, it is anticipated that the dynamical convergence can be efficient through designing a software-defined and re-configured radio access network architecture, so that the communication, data caching, and computing can be dynamically re-organized.

\noindent\textbf{Hardware Constraints on Edge AI:} The effectiveness of edge AI algorithms often relies on the ever-increasing computing power of devices. However, 6G devices will be pervasive then they are with limited communication resource, computing power, and storage suitable for the internet of things applications. Applying the existing AI algorithms may result in the degradation of performance. To address this issue, novel and lightweight edge AI algorithms for 6G devices should be developed.

\subsection{New Core Network: A Three-Dimensional Architecture And Intelligent Mobility Management}	

\noindent In the past, terrestrial mobile system and satellite mobile system are independent, and both are two-dimensional network architectures. From technology perspective, the 6G will integrates terrestrial wireless mobile communication, medium and low orbit satellite mobile communication and short distance direct communication technologies. At the same time, 6G will integrate communication and computing, navigation, perception, and other new technologies. 6G will thus establish a new three-dimensional(3D) core network architecture which can integrate those systems, and support global ubiquitous coverage of high-speed mobile communications, including communications in the air, in space, on the earth and over the sea, by leveraging intelligent mobility management and control methods.

Although this 3D core network architecture can break through traditional coverage limitation and eventually form an unprecedented universal coverage, there meet following challenges to be solved before it is really available.

\noindent\textbf{New Three-dimensional Core Network:} The 6G core network will be a three-dimensional network being composed of network nodes of terrestrial system and network nodes of satellite system. As part of the converged network, a satellite node will be a part of a serving core network through integrating some network functions into the satellite, such as Access and Mobility Management Function (AMF), Session Management Function (SMF), and User Plane Function (UPF). UE can establish the multiple Packet Data Unit (PDU) sessions through the terrestrial node or/and the satellite node by unified protocol. The 6G network is the service-driven intelligent network, and it can allocate the network resources dynamically according to the service requirements and user's priority. The 6G core network will dynamically establish the connections of the user plane following the global view of the network located on the terrestrial system and the satellite system.

\noindent\textbf{Intelligent Mobility Management:} The mobility management of 5G and pre 5G did not take network node movement into account. The fast moving satellites will bring new challenges to 6G, the mobility scenarios of 6G will become more complicated. The inter-satellite links change while satellite position changes. The movement will impact the network topology, handover control methods and etc. Fortunately, the trajectory of satellites is known due to the pre-determined ephemeris and should be used as an important input to the 6G mobility management. The moving prediction of a terminal shall also be enabled by AI. For terminals with on-going service connections, the intelligent mobility management should tackle different types of handovers, such as handover between beams, handover between satellites, handover between airplane stations and satellite/earth stations, and handover between satellite and earth stations. The handover decision must consider the trajectory of satellites, the characteristics of users and infrastructures e.g. bandwidth, delay, signal strength of access links against the quality requirement of the on-going services, as well as load balance among satellites and terrestrial systems. The overall design of handover control procedure should consider the different signaling transmission delays of messages traversing the satellite and terrestrial links. For the terminals without connections, how to achieve reasonable tradeoff between paging signaling cost and offered performance is also a challenge.

\noindent\textbf{Efficient network resource utilization:} Based on the global coverage and overall coordination capability, multi-dimensional information should be processed effectively, wireless resources should be allocated uniformly and efficiently, and different networks should be managed and controlled on demand. Though terrestrial network and satellite network have different topologies and different stability in different scenarios, with the service-driven, intelligent, and three-dimensional core network, also with dynamic change between terrestrial and satellite networks will be hidden behind data transmission. Therefore, a user can get transparent access in the integrated network, and flexible roam between terrestrial access technology and satellite access technology.

\noindent\textbf{Protocol interoperability:} In terrestrial network, usually TCP/IP protocol is used. In satellite network, since the communication channel has ultra wide bandwidth, long transmission delay and higher bit error rate, traditional TCP/IP protocol will face low efficiency due to its original design only for computer network. Therefore, TCP/IP protocol needs to be optimized for satellite network, or novel protocols such as delay tolerant network (DTN) protocol may be used. Since different communication protocols may be used for terrestrial network and satellite network, the heterogeneous network may experience multiple conversions between protocols, resulting in formidable challenges for the network integration. Therefore, different protocols should have interoperability, to satisfy seamless transfer between terrestrial network and satellite network.

\section{Roadmap Suggestion of 6G Standardization}
	
\begin{figure}
  \centering
  \renewcommand{\figurename}{Figure}
  \includegraphics[width=8cm]{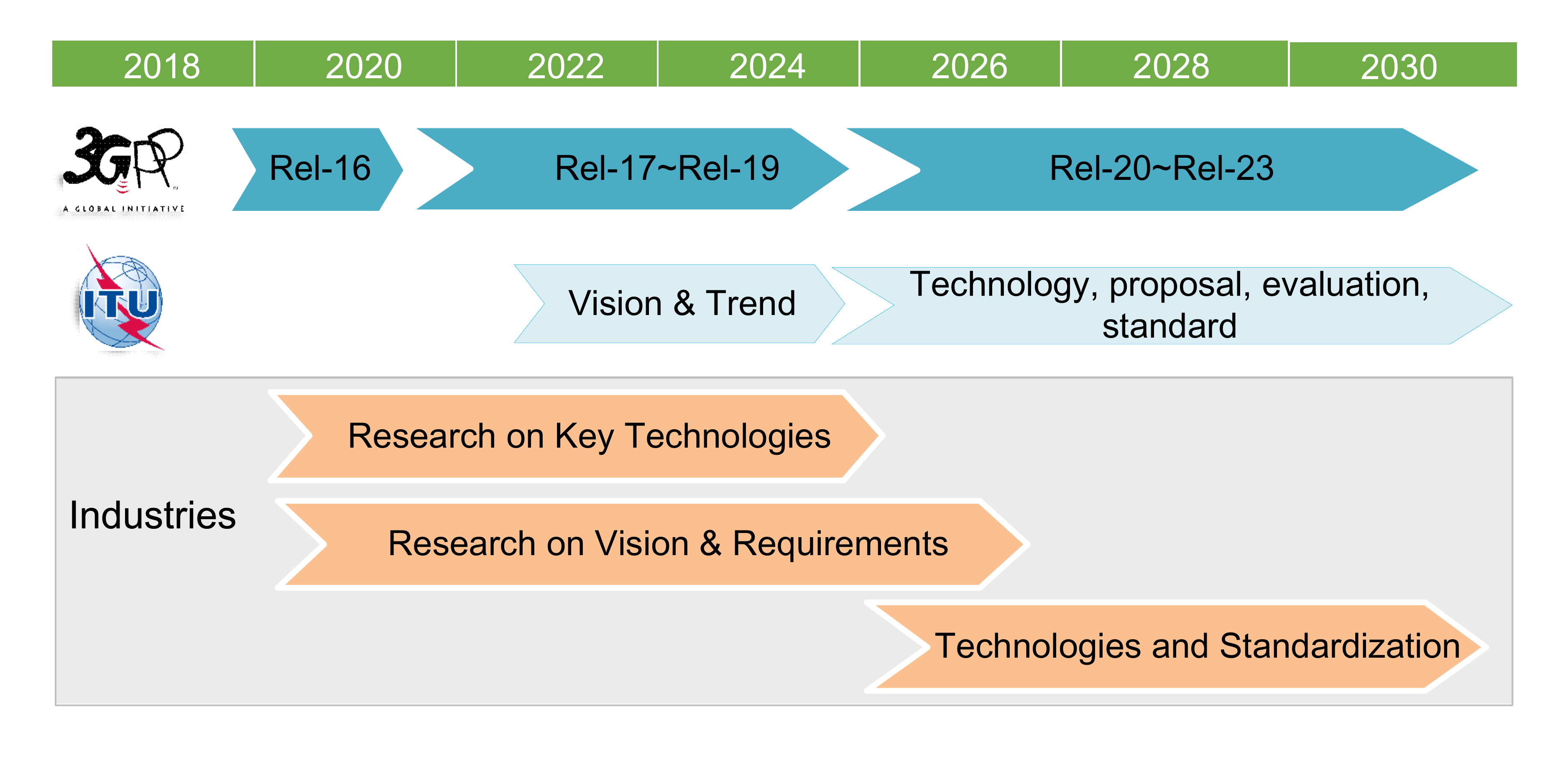}
  \caption{6G Roadmap.}\label{Fig:fig4}
\end{figure}

\noindent An expected roadmap for 6G is shown in Fig. \ref{Fig:fig4}. From the standard organization 3GPP, present release 16 for 5G NR will be finished early 2020, the research on beyond 5G will thus be processed from release 17 to release 19, and research for 6G may be further followed by release 20. From the standard organization ITU, 5G standardization will be issued by the end of 2020, the research on 6G can then be started, with the vision and technology trends being considered first. Simultaneously, from the industry aspect, research on 6G has been started. Key technologies, visions and requirements are being discussed by academics and industries all over the world. It's expected that these researches will continue until converging during the period 2024-2026, the standard work for 6G will then start and will work toward 2030.

In 5G stage, close international cooperation was established to achieve a global standard. For 6G, cooperation should be enhanced further, not only covering present international cooperation from different countries and organizations, but also considering contributions from verticals even at the initial stage of 6G research. In addition, since new material for device and equipment will be challenged toward its theoretical limitation, there should have much closer cooperation among enterprises in the hierarchical industrial chain.

\section{Conclusions}	
\noindent Following the standard optimization and pre-commercial stage of 5G, research on 6G has been under discussion gradually. Though there has not seen any officially agreed opinion on what will be the 6G, as a future novel generation, 6G will no doubt have ten to hundred times higher overall capabilities than that of 5G. From technology perspective, 6G will integrate terrestrial wireless mobile communication, medium and low orbit satellite communication and short distance direct communication technologies, as well as integrate communication, computing, navigation, perception, artificial intelligence and other new technologies. Standard work for 6G will take place around 2025 and 6G pre-commercial network will be available around year 2030. Finally, 6G will support the development of a Ubiquitous Intelligent Mobile Society with intelligent life and industries.

\section*{Acknowledgment}

We would like to thank Dr. Ming Ai and Dr. Hui Xu from Datang Telecom Group and Prof. Yan Shi and Ms Yadan Zheng from Beijing University of Post Telecommunication for their help in preparing related materials, and Prof. Dake LIU from Beijing Institute of Technology for his valuable comments.





%



\bibliography{refer}

%

\begin{IEEEbiography}[{\includegraphics[width=1in,height=1.25in,clip,keepaspectratio]{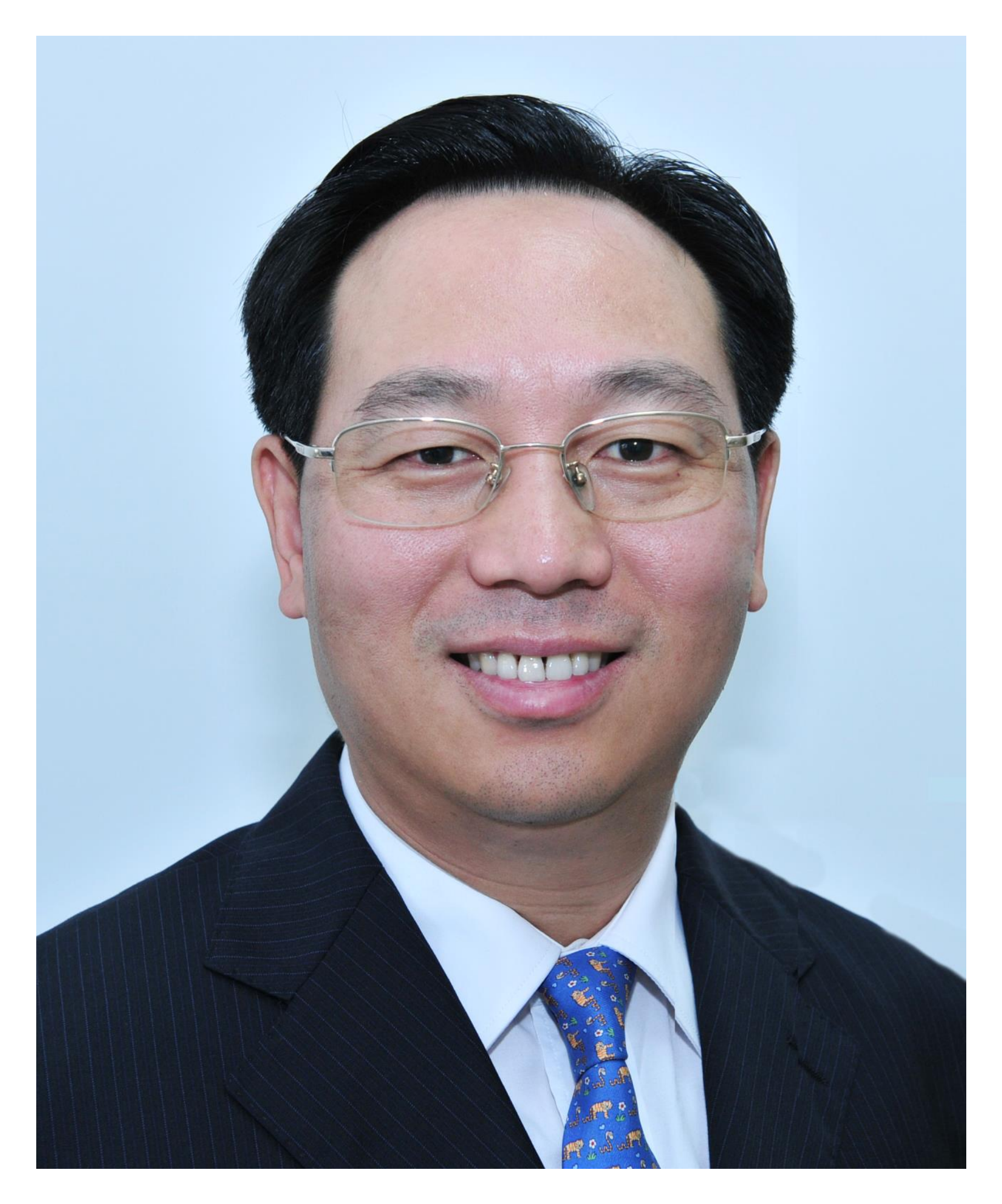}}] {Shanzhi Chen} (SM'04) received the bachelor's degree from Xidian University in 1991 and the Ph.D. degree from the Beijing University of Posts and Telecommunications, China, in 1997. He joined the Datang Telecom Technology and Industry Group and the China Academy of Telecommunication Technology (CATT) in 1994, and has been serving as the EVP of Research and Development since 2008. He is currently the Director of the State Key Laboratory of Wireless Mobile Communications, CATT, where he conducted research and standardization on 4G TD-LTE and 5G. He has authored and co-authored four books [including the well-known textbook Mobility Management: Principle, Technology and Applications (Springer Press)], 17 book chapters, more than 100 journal papers, 50 conference papers, and over 50 patents in these areas. He has contributed to the design, standardization, and development of 4G TD-LTE and 5G mobile communication systems. His current research interests include 5G mobile communications, network architectures, vehicular communication networks, and Internet of Things. He served as a member and a TPC Chair of many international conferences. His achievements have received multiple top awardsand honors by China central government, especially the Grand Prize of the National Award for Scientific and Technological Progress, China, in 2016 (the highest Prize in China). He is the Area Editor of the IEEE Internet of Things, the Editor of the IEEE Network, and the Guest Editor of the IEEE Wireless Communications, the IEEE Communications Magazine, and the IEEE Transactions on Vehicular Technology.
\end{IEEEbiography}

\begin{IEEEbiography}[{\includegraphics[width=1in,height=1.25in,clip,keepaspectratio]{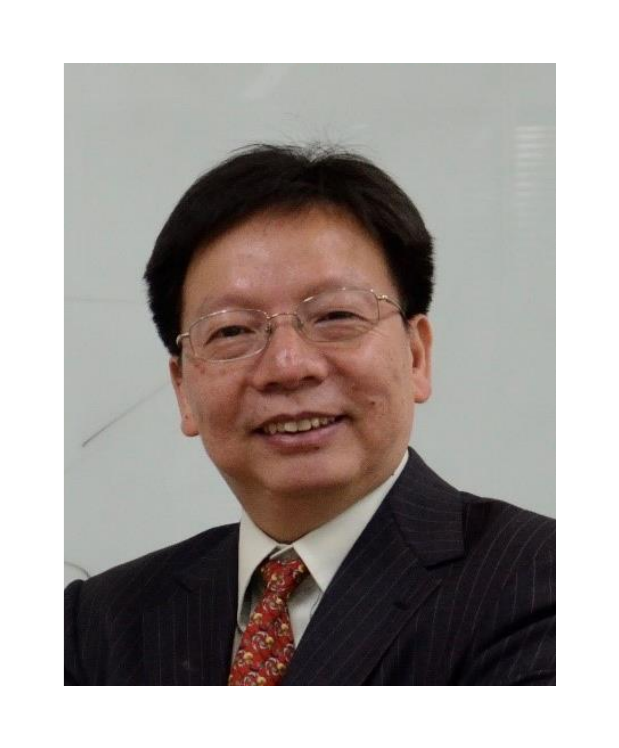}}] {Ying-Chang Liang} (F'11) is currently a Professor with the University of Electronic Science and Technology of China, China, where he leads the Center for Intelligent Networking and Communications and serves as the Deputy Director of the Artificial Intelligence Research Institute. He was a Professor with The University of Sydney, Australia, a Principal Scientist and Technical Advisor with the Institute for Infocomm Research, Singapore, and a Visiting Scholar with Stanford University, USA. His research interests include wireless networking and communications, cognitive radio, symbiotic networks, dynamic spectrum access, the Internet-of-Things, artificial intelligence, and machine learning techniques.

Dr. Liang has been recognized by Thomson Reuters (now Clarivate Analytics) as a Highly Cited Researcher since 2014. He received the Prestigious Engineering Achievement Award from The Institution of Engineers, Singapore, in 2007, the Outstanding Contribution Appreciation Award from the IEEE Standards Association, in 2011, and the Recognition Award from the IEEE Communications Society Technical Committee on Cognitive Networks, in 2018. He is the recipient of numerous paper awards, including the IEEE Jack Neubauer Memorial Award, in 2014, and the IEEE Communications Society APB Outstanding Paper Award, in 2012. He is a Fellow of IEEE, and a foreign member of Academia Europaea.

He is the Founding Editor-in-Chief of the IEEE Journal on Selected Areas in Communications: Cognitive Radio Series, and the Key Founder and now the Editor-in-Chief of the IEEE Transactions on Cognitive Communications and Networking. He is also serving as an Associate Editor-in-Chief for China Communications. He served as a Guest/Associate Editor of the IEEE Transactions on Wireless Communications, the IEEE Journal of Selected Areas in Communications, the IEEE Signal Processing Magazine, the IEEE Transactions on Vehicular Technology, and the IEEE Transactions on Signal and Information Processing over Network. He was also an Associate Editor-in-Chief of the World Scientific Journal on Random Matrices: Theory and Applications. He was a Distinguished Lecturer of the IEEE Communications Society and the IEEE Vehicular Technology Society. He was the Chair of the IEEE Communications Society Technical Committee on Cognitive Networks, and served as the TPC Chair and Executive Co-Chair of the IEEE Globecom'17.

\end{IEEEbiography}

\begin{IEEEbiography}[{\includegraphics[width=1in,height=1.25in,clip,keepaspectratio]{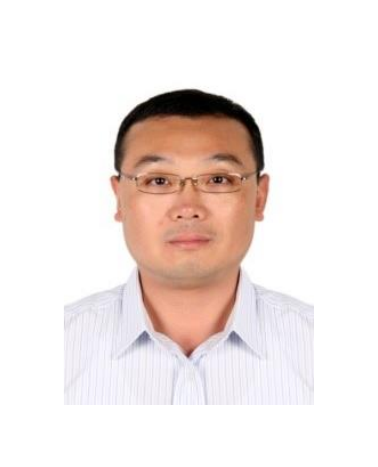}}] {Shaohui Sun} received his Ph.D. from Xidian University, Xi'an, China, in 2003. Form March 2003 to June 2006, he was a postdoctoral research fellow at the Datang Telecom technology and industry group, Beijing, China. From June 2006 to December 2010, he worked at the Datang mobile communications equipment company Ltd., Beijing, where he has been deeply involved in the development and standardization of the Third-Generation Partnership Project Long-Term Evolution (3GPP LTE). Since January 2011, he has been the Chief Technical Officer with Datang Wireless Mobile Innovation Center of the Datang Telecom technology and industry group. In 2019, he joined Datang mobile communications equipment company Ltd. Again, and has served as EVP R\&D. Now, his research area of interest includes advanced technologies related to 5G/6G.
\end{IEEEbiography}

\begin{IEEEbiography}[{\includegraphics[width=1in,height=1.25in,clip,keepaspectratio]{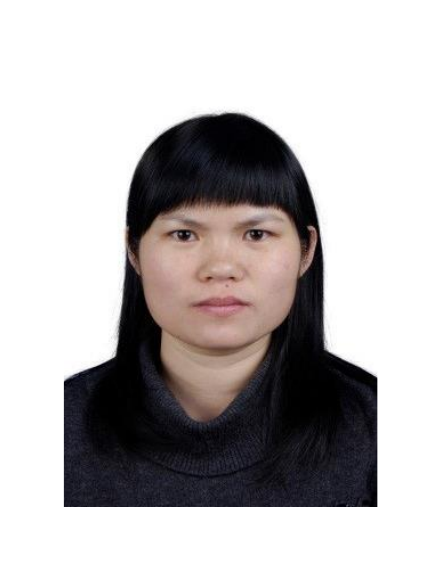}}] {Shaoli Kang}  (kangshaoli@catt.cn) received her Ph.D. degree from Beijing Jiaotong University (BJTU), China, in 2002. From 2000 to 2005, she joined Datang Telecom Technology Industry Group focusing on R\&D of TD-SCDMA. From 2005 to 2007, she acted as a research fellow in the CCSR at the University of Surrey (UniS), UK. Since 2007, she has continued with Datang Telecom Group doing research on 4G and beyond. In recent years she has been working on 5G and presided over the 863 5G project ''5G modulation and coding''. She is active in organizations like CCSA and IMT-2020(5G) PG.
\end{IEEEbiography}

\begin{IEEEbiography}[{\includegraphics[width=1in,height=1.25in,clip,keepaspectratio]{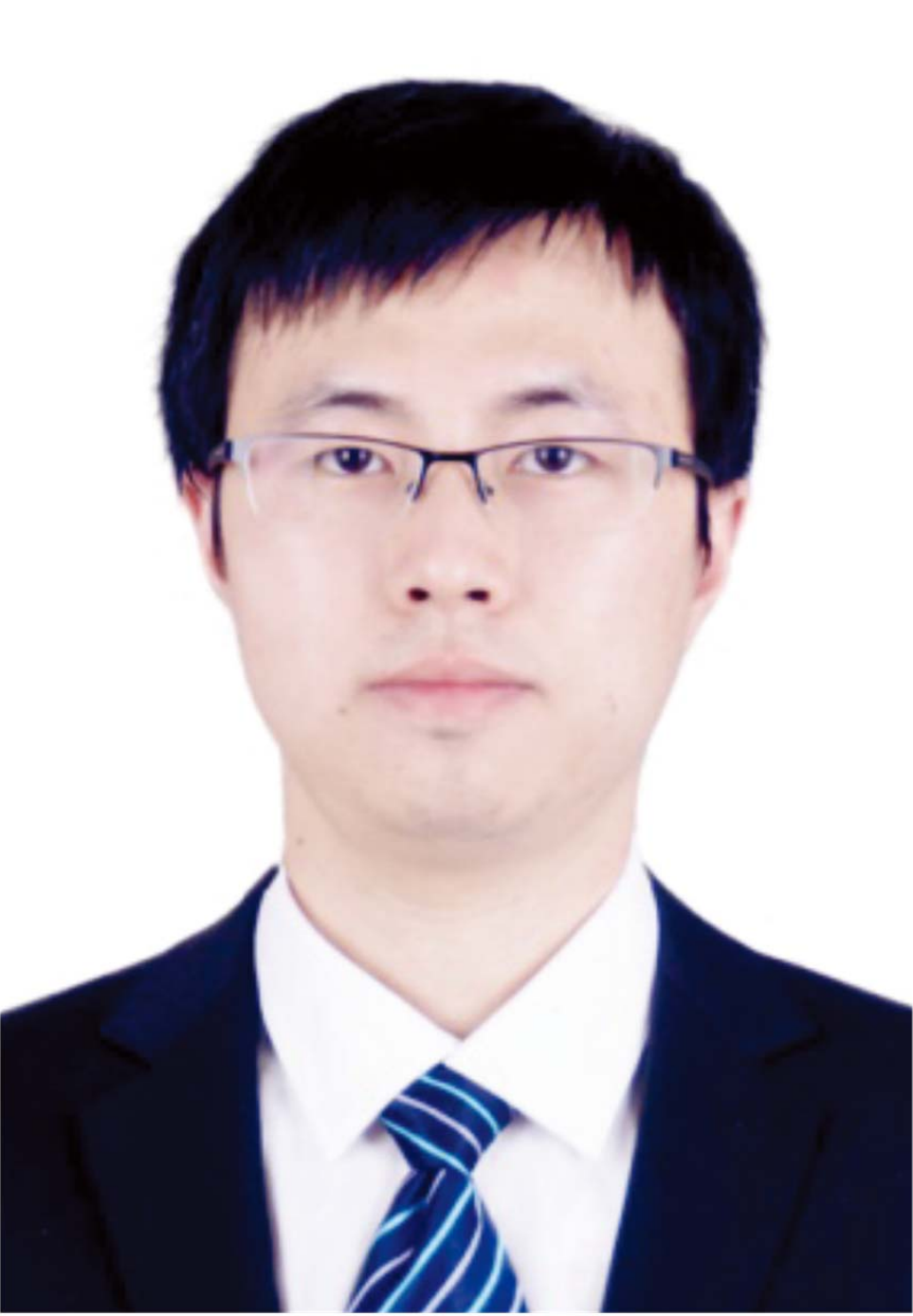}}] {Wenchi Cheng} (M'14-SM'18) received his B.S. and Ph.D. degrees in telecommunication engineering from Xidian University, Xi'an, China, in 2008 and 2013, respectively, where he is an associate professor. He joined the Department of Telecommunication Engineering, Xidian University, in 2013, as an assistant professor. He was a visiting scholar with Networking and Information Systems Laboratory, Department of Electrical and Computer Engineering, Texas A\&M University, College Station, TX, USA, from 2010 to 2011. His current research interests include 5G wireless networks and orbital-angular-momentum based wireless communications. He has published more than 80 international journal and conference papers in IEEE Journal on Selected Areas in Communications, IEEE Magazines, IEEE INFOCOM, GLOBECOM, and ICC, etc. He received URSI Young Scientist Award (2019), the Young Elite Scientist Award of CAST (2016-2018), the Best Dissertation (Rank 1) of China Institute of Communications, the Best Paper Award for IEEE ICCC 2018, the Best Paper Nomination for IEEE GLOBECOM 2014, and the Outstanding Contribution Award for Xidian University. He has served or been serving as the associate editor for IEEE Communications Letters and IEEE Access, the IoT Session Chair for IEEE 5G Roadmap, the Wireless Communications Symposium Co-Chair for IEEE GLOBECOM 2020, the Publicity Chair for IEEE ICC 2019, the Next Generation Networks Symposium Chair for IEEE ICCC 2019, the Workshop Chair for IEEE ICC 2019 Workshop on Intelligent Wireless Emergency Communications Networks, the Workshop Chair for IEEE ICCC 2017 Workshop on Internet of things, the Workshop Chair for IEEE GLOBECOM 2019 Workshop on Intelligent Wireless Emergency Communications Networks.
\end{IEEEbiography}

\begin{IEEEbiography}[{\includegraphics[width=1in,height=1.25in,clip,keepaspectratio]{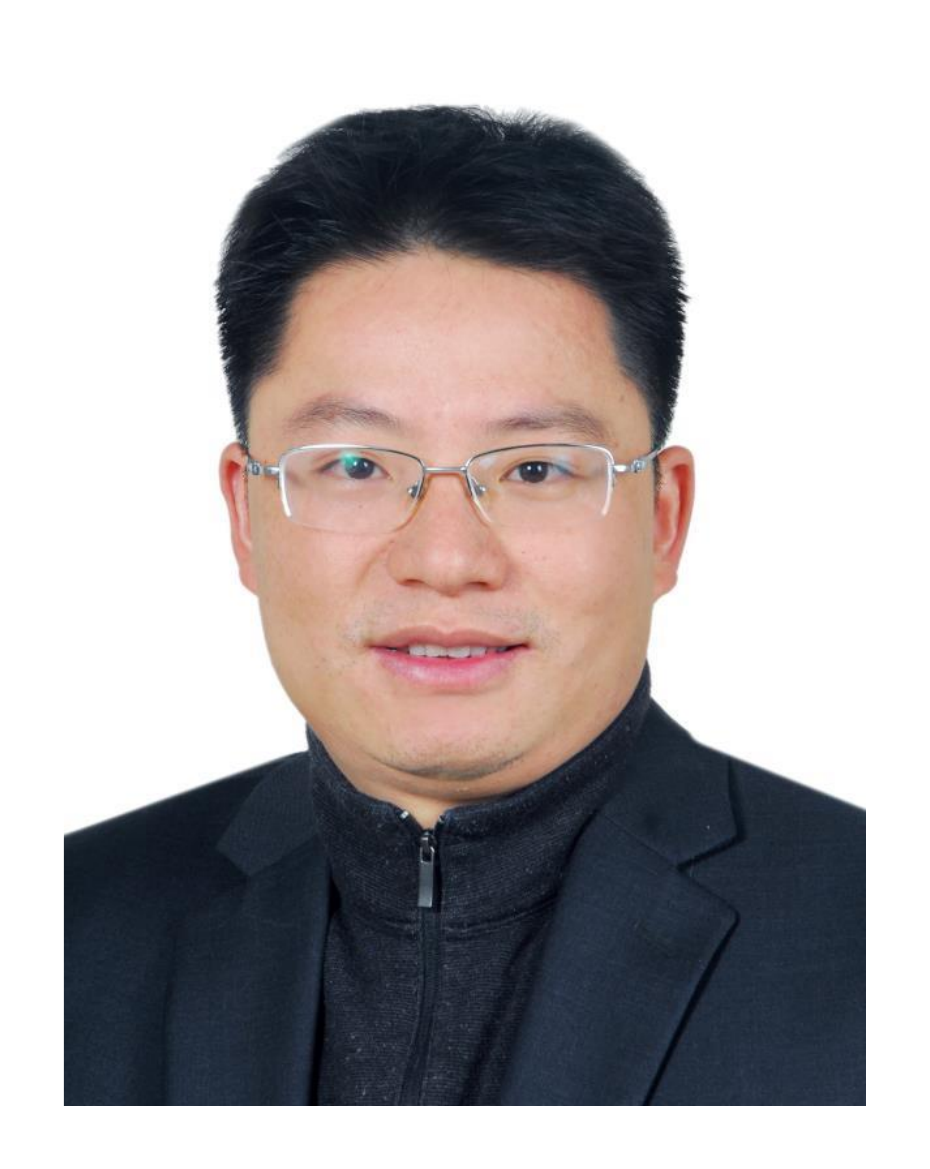}}] {Mugen Peng} (M'05-SM'11) received the Ph.D. degree in communication and information systems from the Beijing University of Posts and Telecommunications (BUPT), Beijing, China, in 2005. He has been a Full Professor with the School of Information and Communication Engineering, BUPT since 2012. In 2014, he was also an Academic Visiting Fellow with Princeton University, USA. He leads a Research Group focusing on wireless transmission and networking technologies in BUPT. He has authored and coauthored over 100 refereed IEEE journal papers and over 300 conference proceeding papers. His main research areas include wireless communication theory, radio signal processing, cooperative communication, self-organization networking, heterogeneous networking, cloud communication, and Internet of Things.

He was a recipient of the 2018 Heinrich Hertz Prize Paper Award, the 2014 IEEE ComSoc AP Outstanding Young Researcher Award, and the Best Paper Award in the JCN 2016, IEEE WCNC 2015, IEEE GameNets 2014, IEEE CIT 2014, ICCTA 2011, IC-BNMT 2010, and IET CCWMC 2009. He is currently or have been on the Editorial/Associate Editorial Board of IEEE Communications Magazine, IEEE ACCESS, the IEEE INTERNET OF THINGS JOURNAL, IET Communications, and China Communications.
\end{IEEEbiography}






\end{document}